\documentclass[conference]{IEEEtran}
\IEEEoverridecommandlockouts
\usepackage{cite}
\usepackage{amsmath,amssymb,amsfonts}
\usepackage{algorithmic}
\usepackage{graphicx}
\usepackage{textcomp}
\usepackage{xcolor}

\usepackage[printonlyused]{acronym}
\usepackage[caption=false,font=normalsize,labelfont=sf,textfont=sf]{subfig}
\usepackage{url}
\usepackage{booktabs}   
\usepackage{array}      
\usepackage{multirow}  
\usepackage{caption}    

\newcommand{\new}[1]{#1}

\def\BibTeX{{\rm B\kern-.05em{\sc i\kern-.025em b}\kern-.08em
    T\kern-.1667em\lower.7ex\hbox{E}\kern-.125emX}}

\acrodef{InfoVis}{Information Visualization}
\acrodef{PD}{participatory design}
\acrodef{MTG}{Magic: The Gathering}
\acrodef{TCG}{trading card game}
\acrodef{AI}{artificial intelligence}
\acrodef{HCI}{human-computer interaction}
    
\begin{document}

\title{Building an MTG Data Dashboard:\\Initial Design}

\author{Blind for Review}

\author{\IEEEauthorblockN{Tomás Alves\textsuperscript{3}, João Moreira\textsuperscript{1,2,*}}\IEEEauthorblockA{\textsuperscript{1}\textit{COPELABS, INESC INOV-Lab, Lusófona University, Lisbon, Portugal}\\\textsuperscript{2}\textit{INESC-ID, Lisboa, Portugal}\\\textsuperscript{3}\textit{Business Research Unit, ISCTE-IUL, Lisbon, Portugal}\\{\textsuperscript{*} Email: joao.moreira@ulusofona.pt}}}

\maketitle

\begin{abstract}
This paper presents the initial stages of a design study aimed at developing a dashboard to visualize gameplay data of the Commander format from Magic: The Gathering.
We conducted a user-task analysis to identify requirements for a data visualization dashboard tailored to the Commander format.
Afterwards, we proposed a design for the dashboard leveraging visualizations to address players' needs and pain points for typical data analysis tasks in the context domain.
Then, we followed-up with a structured user test to evaluate players' comprehension and preferences of data visualizations.
Results show that players prioritize contextually relevant, outcome-driven metrics over peripheral ones, and that canonical charts like heatmaps and line charts support higher comprehension than complex ones such as scatterplots or icicle plots.
Our findings also highlight the importance of localized views, user customization, and progressive disclosure, emphasizing that adaptability and contextual relevance are as essential as accuracy in effective dashboard design.
Our study contributes practical design guidelines for data visualization in gaming contexts and highlights broader implications for engagement-driven dashboards.
\end{abstract}

\begin{IEEEkeywords}
human-centered computing, information visualization, visualization design, evaluation methods
\end{IEEEkeywords}

\section{Introduction}

Recent research in modern board and card games focuses on leveraging data-driven methods, \ac{AI} models, and \ac{InfoVis} techniques to support decision-making, strategy development, and player understanding.
Among these games, \ac{MTG} stands as the progenitor of the wider \ac{TCG} genre as it is known today and the world’s largest tabletop collectible card game~\cite{ohm2023magic,zhu2023intelligence}.
Specifically for \ac{MTG}, some studies focus on generalized card representations, where multi-modal embeddings combine textual, numeric, visual, and usage-based features to capture card properties across expansions~\cite{bertram2024learning}.
Other researchers tackle how game complexity has evolved through the years~\cite{magruder2022conservative} or card synergies~\cite{xue2025cardsynergy} using data visualizations.
Another major area of research examines how \ac{AI} agents and prediction models help drafting and human decision-making~\cite{ward2021ai,bertram2021predicting,zhu2023intelligence,alvin2021toward}.
Traditional drafting is a complex process requiring players to balance immediate card value against long-term deck coherence.
Beyond mechanics and data, perceptual and aesthetic factors also play a significant role.
For instance, Kallabis et al.~\cite{kallabis2025deceptive} showed that visual styling can bias how players perceive card strength, even when objective properties remain constant.
Finally, other studies focus on the gamer culture of \ac{MTG}~\cite{trammell2010magic,falcao2021conservatism}, predicting personality traits and color preferences~\cite{cui2025identity}, or the cultural aspects of designing cards~\cite{fornazari2016gender,zanescu2022designing}.

Together, these studies highlight an emerging research space at the intersection of AI, visualization, and trading card games.
However, we identified two critical research gaps in \ac{MTG} related research.
First, \new{\textbf{there is a lack of research on \ac{MTG}'s most popular format, Commander}}.
This format has become especially popular due to its large deck size, singleton rules, and focus on synergy-driven gameplay.
This richness makes deck construction and strategy development both rewarding and challenging, as players must navigate vast card pools, hidden synergies, and shifting metagames.
Second, although community-driven resources and APIs (e.g., Scryfall \cite{scryfallScryfallMagic} and MTGJSON \cite{mtgjsonMTGJSON}) and deck-building websites (e.g., Archidekt \cite{archidektDeckBuilder} or Moxfield \cite{moxfieldMoxfield}) provide comprehensive card data or data visualizations to track deck metrics, \new{\textbf{there is currently no research on visual analytics for MTG gameplay data}}, leaving players without interactive tools to explore strategic patterns or deck performance.
\new{Therefore, we believe that a visualization dashboard enabling players to explore the game's rich dataset will enhance the visualization community by demonstrating effective visual idioms and task designs, while also improving players' awareness of gameplay mechanics in this complex game.}

In this paper, we report on the initial stages of a design study to develop a visualization dashboard for Commander.
Our approach draws from design studies~\cite{sedlmair2012design} in visualization research, focusing on supporting players in understanding synergies, situating their decks within broader strategic contexts, and assessing their performance against other players.
Specifically, we contribute \new{to \ac{HCI} evaluation methods} with:
\begin{itemize}
    \item A user-task analysis to identify requirements for a data visualization dashboard tailored to the Commander format of \ac{MTG}.
    \item An evaluation of users' comprehension and preferences through a structured user test of data information visualizations.
\end{itemize}
\section{Related Work}

User-centered design~\cite{norman1986user} has become a well-established framework in \ac{HCI}, emphasizing integration of stakeholder needs into the design process~\cite{fdis20099241,kopec2017livinglab}. Engaging users accelerates prototyping and improves outcome quality~\cite{signoretti2015trip,lindsay2012engaging}. Participatory design (\ac{PD}), a branch of user-centered design, enables active user contribution to identifying design directions~\cite{lee2017steps,frishberg2011interactive}. Evidence shows \ac{PD} increases satisfaction in private sectors~\cite{grissemann2012customer} and drives social innovation publicly~\cite{voorberg2015systematic}. \new{In this section, we explain relevant \ac{HCI} design frameworks motivating our proposal's development, review game visual analytics state-of-the-art, and specify requirements for \ac{MTG}.}

\subsection{Design Studies in Information Visualization}

Design studies extend \ac{PD} methodology by involving users throughout design~\cite{lam2017bridging}. Sedlmair et al.~\cite{sedlmair2012design} define a design study as analyzing real-world problems, designing visualization systems, validating designs, and reflecting on lessons learned. Applied across domains~\cite{goodwin2016constraint,nobre2018lineage}, close researcher-stakeholder collaboration strengthens trust and enables joint exploration of visualization opportunities~\cite{goodwin2016constraint,kerzner2018framework}.

Recent work demonstrates the value of user-centered, multi-view, and explainable approaches~\cite{wang2022extending,ruan2024violet,sarigai2025dciwebmapper,kandel2024pd}: dciWebMapper explores geospatial data through interactive views; PD-Insighter monitors patient behaviors via temporal/spatial visualizations; Violet interprets quantum neural networks; Wang et al. apply explainable AI to drug repurposing. These studies highlight design principles applicable to trading card game dashboards: multi-view coordination, exploratory analysis support, complex relationship explanation, and grounding interfaces in domain expertise—guiding our \ac{MTG} requirement identification and visualization proposals.

\new{\subsection{Data Visualizations for Games}}

Game analytics visualization has grown over the past decade as competitive markets demand understanding player behavior for improved game quality~\cite{Wallner2014,Wallner2015,Wallner2018}. Internet-enabled devices accumulate vast player-game interaction data, driving the emergence of game analytics—discoveries communicated through visualization techniques~\cite{Bowman2012}.

Common methods include charts, heatmaps, and scatterplots for statistical analysis, though sequential player behavior information is often neglected~\cite{Teng2024}. Heatmaps visualize metrics like deaths and spatial distribution~\cite{Kriglstein2014,Moura2011}. Advanced approaches include node-edge graphs~\cite{Teng2023}, hexbin maps~\cite{Wallner2020}, animated MOBA maps~\cite{Goncalves2018}, and integrated systems like PLATO~\cite{Wallner2014}. Challenges persist: dynamic high-dimensional data, scalability for individual traces, and stakeholder customization needs~\cite{Wallner2018,Teng2023}. For example, research suggests heatmaps excel at hotspot detection while clusters better reveal variable relationships~\cite{Kriglstein2014}.

\subsection{Data Visualizations for Magic: The Gathering}

Current literature offers limited evidence applying data visualization to card synergies, gameplay, or collections. \new{Existing \ac{TCG} literature lacks studies focused on visualizing gameplay data.} Nevertheless, online resources provide analytical tools. Vincenzo Galante's Looker dashboard~\cite{githubGitHubVincenzoGalantemagicthegathering} uses bar/pie charts for yearly releases and color distribution. Similar explorations employ comparable conventions~\cite{beran}. Archidekt~\footnote{https://archidekt.com} uses pie charts for color counts and bar charts for mana-value distributions.

\begin{figure*}[] \centering \subfloat[Deck dashboard.]{\includegraphics[height=0.2\textheight]{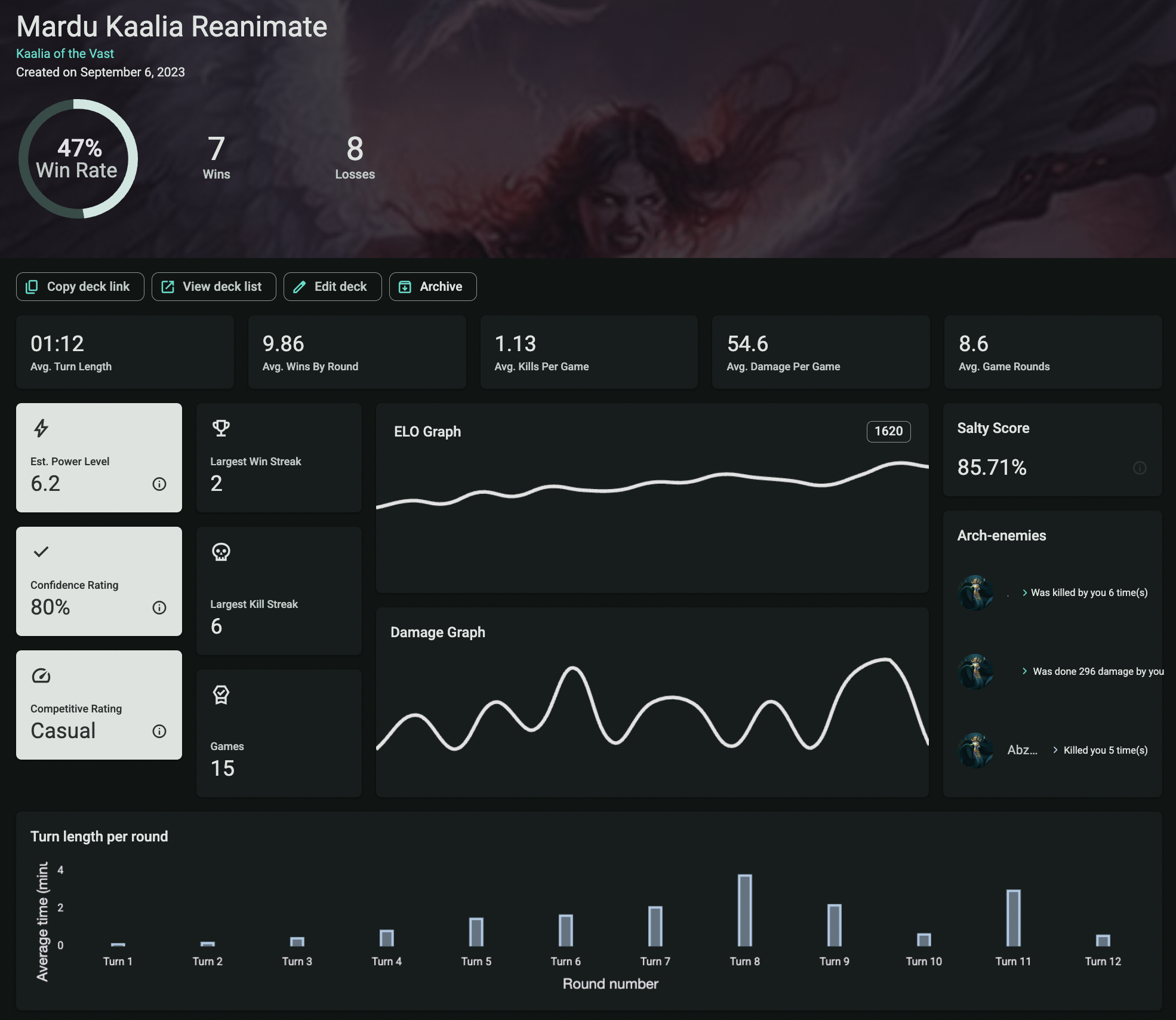} \label{fig:deck-playgroup-gg}} \hfil \subfloat[Player dashboard.]{\includegraphics[height=0.2\textheight]{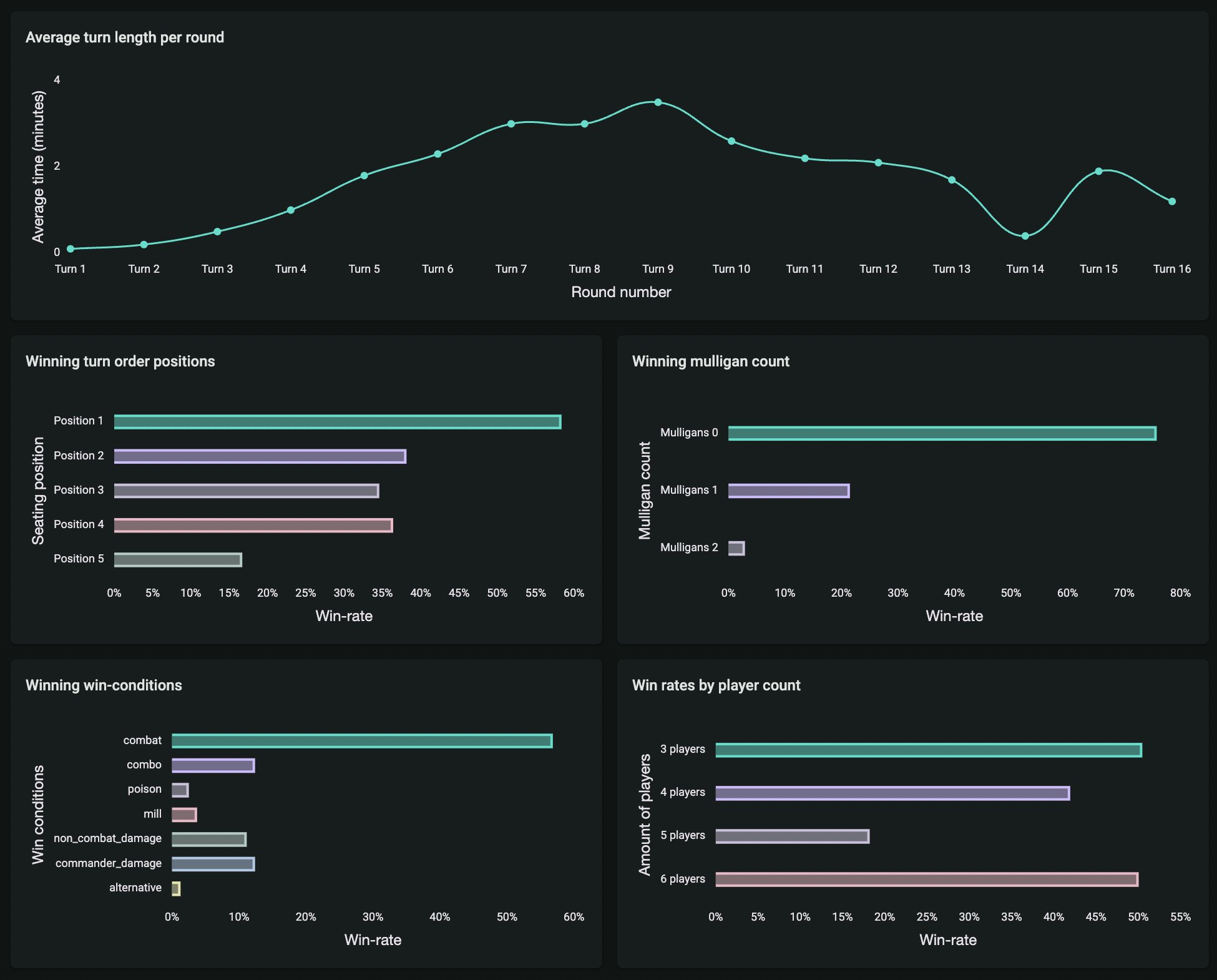} \label{fig:player-playgroup-gg}} \hfil \subfloat[Playgroup dashboard.]{\includegraphics[height=0.2\textheight]{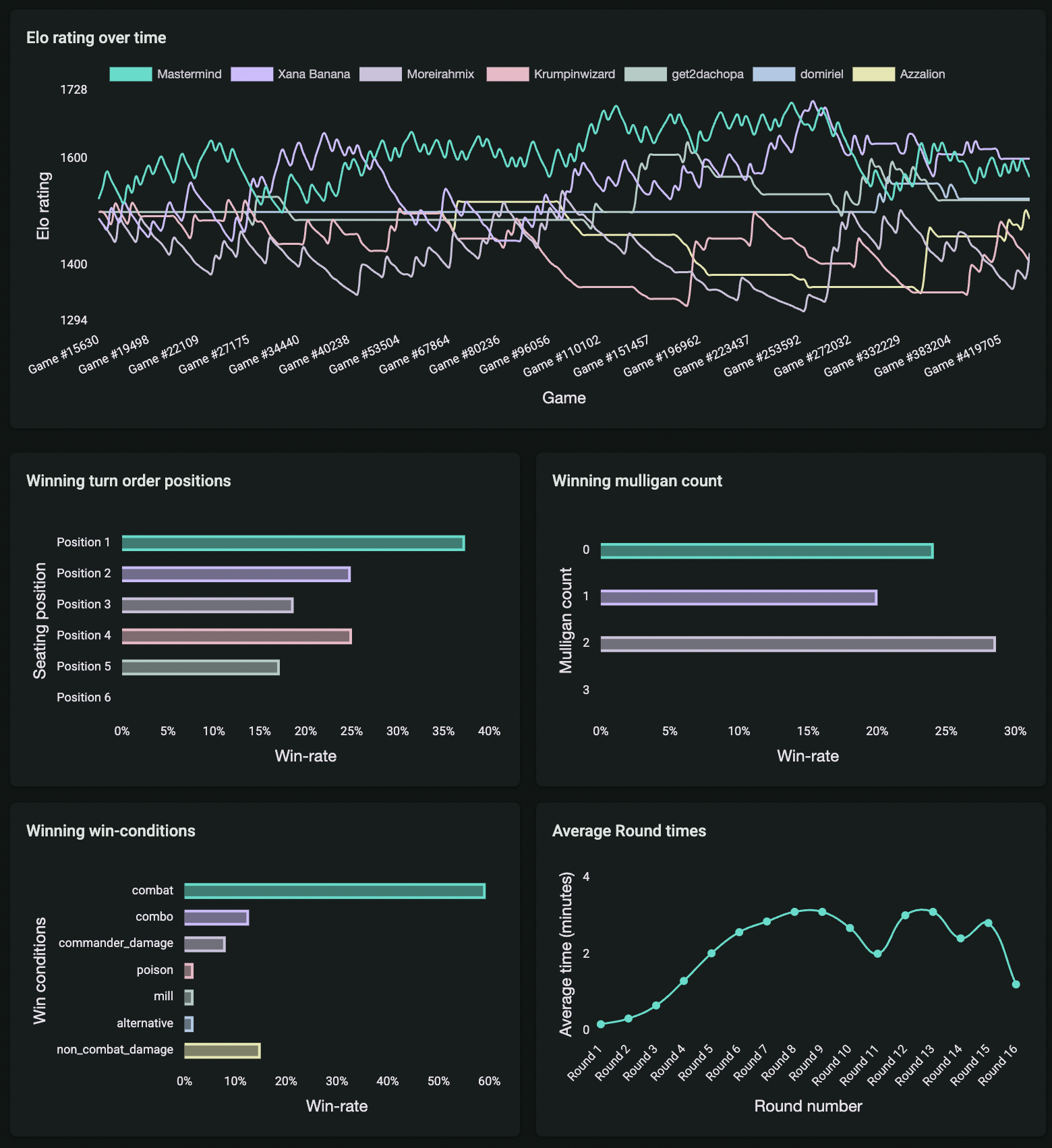} \label{fig:playgroup-playgroup-gg}} \caption{Playgroup.gg dashboards \cite{playgroug} featuring game/deck stats (win rates, turn lengths, commander usage) and player metrics (performance history, social stats).} \label{fig:playgroup-gg} \end{figure*}

Two resources align with our objectives. Pacheco's Commander dashboard~\cite{githubGitHubGabipachecoMTGCommanderAnalysis} analyzes 53 matches between four players (June 2023–August 2024), capturing deck usage, performance, and win conditions via arc diagrams, area charts, and scatterplots. Limitations include custom-only access lacking interactivity and insight generation beyond explicit visuals. Playgroup.gg~\footnote{https://playgroup.gg} tracks Commander games through comprehensive visualization features including game/deck stats (win rates, turn lengths, commander usage) and player dashboards (most-played decks, historical performance, rivalries)—(\ref{fig:playgroup-gg}). Though showcasing high data diversity, charts fragment across pages, hindering integration/filtering. Color palettes impede mark distinction, and finding custom granular stats remains challenging.

These resources provide reference points for developing our Commander dashboard, highlighting common tasks, data dimensions, and chart types. However, limitations reveal improvement opportunities: \new{(i)} most lack peer-reviewed support, being community-developed rather than formally studied; \new{(ii)} focus on descriptive statistics rather than comparative/exploratory analysis restricts pattern identification; \new{(iii)} reliance on color differentiation creates accessibility issues; \new{(iv)} limited interactivity prevents customization; and finally, fragmented visual components impede holistic sensemaking. Our subsequent methodology addresses these issues while catering to player needs.
\section{Methodology}

Sedlmair et al.'s Nine-Stage Framework~\cite{sedlmair2012design} provides a structured methodology for conducting design studies in visualization research, emphasizing iterative collaboration with domain experts. Stages include: \textbf{learn} (domain knowledge acquisition), \textbf{winnow} (problem prioritization), \textbf{cast} (problem abstraction), \textbf{discover} (solution exploration), \textbf{design} (solution refinement), \textbf{implement} (system development), \textbf{deploy} (real-world delivery), \textbf{reflect} (analysis and lessons learned), and \textbf{write} (documentation). The framework emphasizes iterative, overlapping rather than strictly linear progression.

Our study followed stages from \textbf{learn} to \textbf{design}, focusing on understanding players' needs and pain points regarding current data collection/analysis tools and identifying appropriate visualizations for Commander gameplay insights.\new{We conclude with \textbf{reflect}, discussing findings relative to the design phase and outlining future implementation/deployment pathways.} Throughout, we balanced usability with analytical depth, ensuring actionable, context-sensitive insights remained accessible across varying player engagement, technical skill, and visualization literacy levels. This study forms the foundation for subsequent dashboard design. Experimental materials, code, and results are available at \footnote{\url{https://osf.io/rsz9g/overview?view_only=e50cc8aff82e47c580bb422633599487}}.
\section{Precondition Phase}
\new{The first phase of the design study methodology by Sedlmair et al. \cite{sedlmair2012design} is the precondition phase, which comprises three stages: learn, winnow, and cast.}

\subsection{Learn}
As we mentioned, we focus on Commander \cite{wizardsCommanderFormat}  (formerly known as Elder Dragon Highlander or EDH), one of the most popular and distinctive formats in \ac{MTG}.
Unlike competitive one-on-one formats, Commander emphasizes creativity, social interaction, and long-term strategy, making it particularly appealing to casual and community-oriented players.
Each deck contains exactly 100 cards, and with the exception of basic lands, no duplicates are allowed. Central to the format is the choice of a Commander, which is usually a legendary creature or planeswalker, that defines the deck's color identity through its mana symbols and shapes its overall strategy.
Games are most often played with four players, each starting at 40 life points, which is double the usual starting total.
The Commander itself is always available to cast from the Command Zone and can be replayed with an increasing two-mana tax each time it is removed.
In addition to regular win conditions, the format introduces a unique rule called Commander Damage, where a player loses the game if they receive 21 combat damage from a single Commander over the course of the game.
Commander is strongly defined by its synergy-driven deck-building.
The singleton rule and the enormous card pool of more than 20,000 unique cards foster highly diverse and often thematic decks.
Social dynamics, such as table politics, threat perception, and temporary alliances, further distinguish the format from competitive play.
This combination of diversity, strategic depth, and strong community engagement has made Commander a cornerstone of modern \ac{MTG}.

\subsection{Winnow}

After identifying the problem domain, we continued the design study with the \textbf{winnow} stage.
During this phase, we engaged in initial discussions to identify potential collaborators within the \ac{MTG} community who had a strong alignment between their domain needs and our research objectives.
The Commander format helped raise an ecosystem of deck-building platforms, online discussions, and data-driven tools, making it a particularly rich case study for research in visualization and analytics.
Among these tools, Gauntlet \cite{gauntletappGauntlet}, Playgroup.gg \cite{playgroug}, Dragon Counter \cite{dragoncounterDragonCounter}, LifeLinked \cite{lifecounterLifeCounter}, MTG Life Counter: Lotus \cite{lifecounterLifeCounter} and Mythic Track \cite{mythictrackTrackYour} stand as the preferred tools to track track life, wins, or more general gameplay history.
We analyzed these tools to understand the main features, data availability, and the potential to host a dashboard for Commander data.
Our analysis shows that most tools do a satisfactory job of tracking generic gameplay data (e.g., damage, game record-keeping), but they either track life totals or game records.
Among them, Playgroup.gg stands out as the exception, being able to track life and other gameplay data seemingly.
Playgroup.gg also offers a rich history of game statistics and static dashboards with data visualizations, allowing players to explore their data and compare it against other decks, players, and playgroups (\ref{fig:playgroup-gg}).
Therefore, we contacted the founders and developers of Playgroup.gg, Maran and Johannes, to establish a collaboration.

\subsection{Cast}
The subsequent \textbf{cast} stage involved formalizing this collaboration, confirming mutual expectations, and committing resources to jointly develop the visualization dashboard.
We identified the front-line analysts as every Playgroup.gg user since they interact directly with gameplay data and would use the resulting dashboard.
Additionally, the Playgroup.gg developers are the gatekeepers, who hold authority over data access and project approval.
Finally, we, the authors, act as \ac{HCI} researchers and translators since we need to (i) abstract the domain problems into a more generic form considering larger-context domain goals, (ii) conduct a user task analysis, and (iii) design and evaluate data visualizations to satisfy user needs related to the domain problems.
Establishing these roles early ensured that we could proceed efficiently and conduct a successful iterative design of the dashboard.
\section{Core Phase}
\new{The second phase of the design study methodology comprises four stages: discover, design, implement, and deploy. In this study, we completed only the first two stages; the remaining stages are reserved for future work.}

\subsection{Discover}

During the \textbf{discover} stage, we focused on understanding which data and insights players were most interested in regarding the Commander format, in order to identify opportunities for visualization.

\subsubsection{User-Task Analysis}

We conducted a user-task analysis using a structured questionnaire designed to capture both the strengths of Playgroup.gg's current practices and the challenges users encountered when exploring its existing static dashboards.
\new{Although Sedlmair et al. mention interviews as a viable method \cite{sedlmair2012design} and that interviews generate better insights than surveys \cite{Zunaed2021}, we decided to retain the questionnaires. Questionnaires allow large amounts of information to be summarized efficiently \cite{Reid1988Questionnaires}, reach much larger populations \cite{Zunaed2021}, and generate data that can be easily quantified and statistically analyzed \cite{Reid1988Questionnaires}.}
The questionnaire was organized into four main sections to assess users' interests and needs concerning: (i) gameplay data, (ii) relationships between different attributes, (iii) visual analytical tasks, and (iv) data comparison contexts (e.g., comparisons between a player's decks, within a specific playgroup, or across all users).
Each item was rated on a 5-point Likert scale, ranging from \textit{very uninteresting} (-2) to \textit{very interesting} (2). 
We also collected demographic data to understand our sample and assess its inner diversity.

\subsubsection{Results}

The questionnaire was distributed via Reddit, Discord servers focused on \ac{MTG}, and Playgroup.gg's webpage to ensure participant familiarity with the game domain. Results were analyzed using descriptive statistics~\cite{Nick2007,Gupta2019}. Of 68 participants, 58 (85.29\%) were current Playgroup.gg users.

Players showed the highest interest in performance statistics: win rate (M = 1.62, SD = 0.69) and wins versus losses (M = 1.57, SD = 0.70), followed by deck/player rankings (M = 1.40, SD = 0.88). Gameplay event metrics received moderate interest: damage/healing (M = 1.04, SD = 1.16), rounds (M = 0.99, SD = 0.97), and starting position (M = 0.87, SD = 1.13). Less common statistics ranked lower: estimated power level (M = 0.82), kill frequency (M = 0.76), ELO (M = 0.71). Peripheral/chance-driven metrics showed little to negative interest: priority changes (M = -0.13), dice rolls (M = -0.38), and pause duration (M = -0.63). Overall, users prioritize outcomes and performance metrics over auxiliary or stochastic elements.

Relationships linking game outcomes to key variables generated the strongest interest: win rate by commander (M = 1.71, SD = 0.62), wins by commander (M = 1.49), and deck/player rank by commander (M = 1.43). Moderately interesting included win rate by color identity (M = 1.28), starting position (M = 1.22), and kill frequency by commander (M = 1.21). Marginal relationships received M approximately 0.5–0.9; peripheral factors like priority changes and salt score (informal frustration metric) received M less than or equal to 0. Findings indicate users prioritize analyses revealing how commanders, colors, and performance metrics influence outcomes, while marginal variables show limited interest.

Dashboard view preferences favored Playgroup view (members of specific playgroup; M = 1.32, SD = 0.95), followed by User view (individual player's decks; M = 1.07, SD = 1.20). Global view (all playgroups) was considerably less interesting (M = 0.21, SD = 1.31), suggesting players value contextually relevant comparisons over broad analyses—aligning with Commander's casual, social format. For insights, users prioritized strategically actionable patterns: games played/won/kills per commander (M = 1.22, SD = 1.08). Moderate interest existed for color/commander popularity (M = 0.97); turn time, damage per turn, and damage type distributions received M = 0.46–0.71.

Qualitative analysis revealed additional suggested metrics: archenemy tracking, average turns to cast commander, commander casts per game, personal skill trends, and cross-playgroup/comparative skill analyses. Overall, players value strategically actionable, turn-sensitive, and contextually relevant metrics preferring playgroup/deck-contextual views seeking flexibility to track long-term trends, deck evolution, and analyses connecting to game outcomes.

\new{High-level objectives were established: comparative analysis (\textbf{T1}), relationship mapping (\textbf{T2}), trend identification (\textbf{T3}), and hierarchical evaluation (\textbf{T4}). These shaped the subsequent methodology.}

\subsection{Design}

\new{During the design stage, we translated insights from the discovery phase into concrete visualization solutions for Playgroup.gg, beginning the development of our mock design implementation.}
This process involved generating and validating data abstractions, visual encodings, and interaction mechanisms that could effectively communicate gameplay and strategic information to players.
Before narrowing the proposal space in consultation with the Playgroup.gg founders, we iteratively explored multiple design alternatives, including different chart types, filtering options, and comparison contexts (e.g., individual decks, playgroup-wide, or global views).
We ran two rapid prototyping workshops to create sketches and mockups of an interactive dashboard to enable quick feedback loops, allowing us to refine visual encodings and interaction features based on domain expert input.

\subsubsection{The Dashboard}

\begin{figure*}[]
    \centering
    \fbox{\includegraphics[width=.7\textwidth]{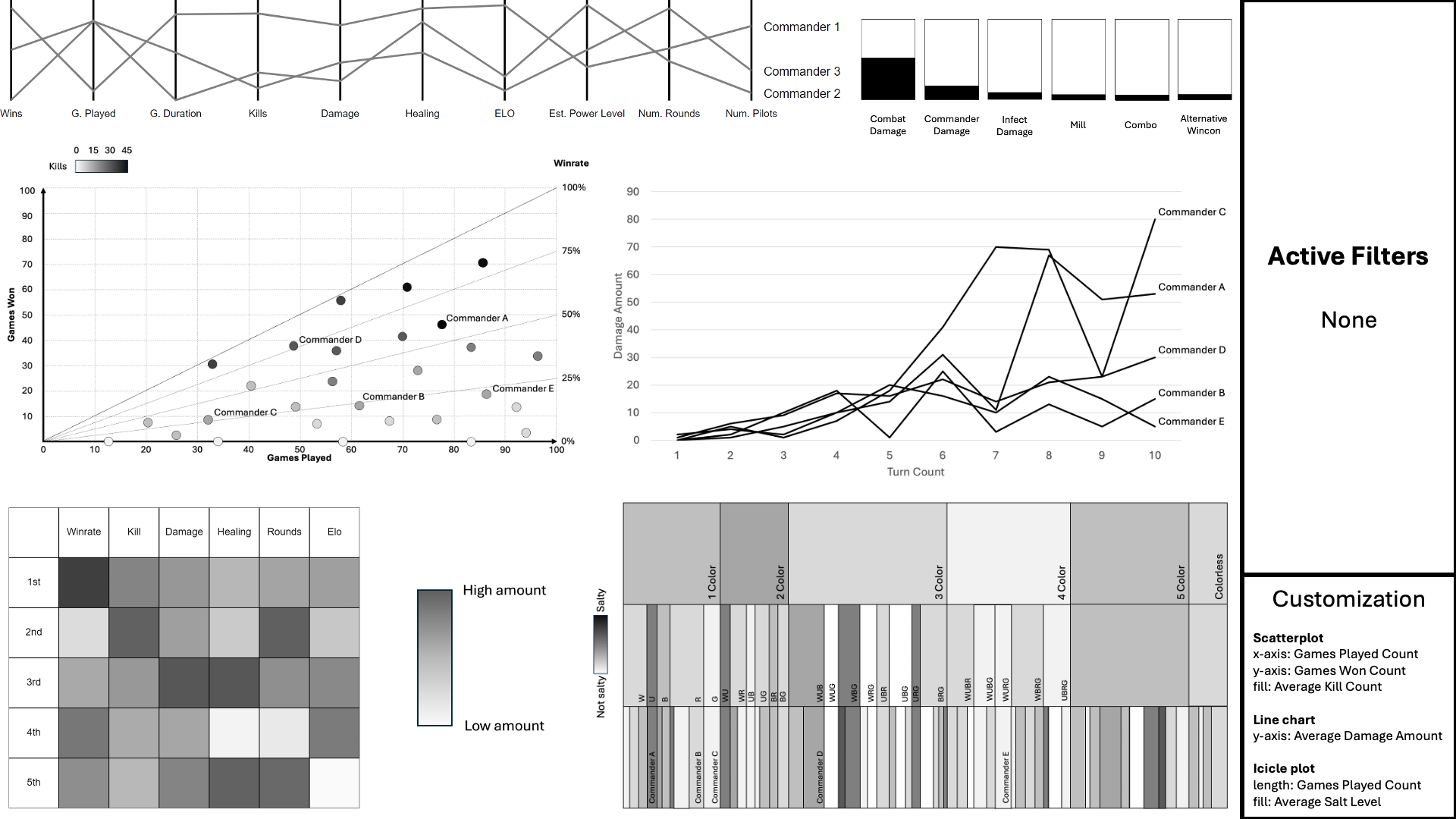}}
    \caption{Mockup of the proposed dashboard for Commander data analysis for Playgroup.gg. \new{The top section features a parallel coordinates plot summarizing key metrics, with bar charts to the right showing the frequency of specific game end states. Below, a scatterplot maps games played against games won, while an adjacent line chart tracks damage dealt per turn by commanders. The lower-left heatmap displays seating positions across multiple metrics. At the bottom-right, an icicle plot encodes deck counts by color identity and number of colors, with brightness indicating player 'saltiness' (a measure of frustration or playstyle intensity). Finally, the dashboard includes panels for active filters and customization instructions.}}
    \label{fig:dashboard-proposal-playgroup}
\end{figure*}

\ref{fig:dashboard-proposal-playgroup} presents the dashboard mockup the authors proposed to the Playgroup.gg team.
The dashboard provides a comprehensive overview of player and deck performance in the Commander format, integrating multiple coordinated visualizations to support analysis at both granular and aggregate levels.
The top section features a parallel coordinates plot \new{(\textbf{T1} and \textbf{T2})} summarizing key metrics such as number of wins and games played, damage, healing, ELO, and estimated deck power, \new{as this type of chart enables users to quickly compare multiple commanders across several dimensions \cite{Blaas2008}.}
To the right, bar charts \new{(\textbf{T1})} illustrate the frequency of specific game end states, facilitating the identification of deck strategies.
Below, a scatterplot \new{(\textbf{T2})} maps games played against games won, with shading encoding average kills, providing a clear view of performance trends, while an adjacent line chart \new{(\textbf{T3})} tracks damage dealt per turn by commanders, supporting temporal analysis.
The lower-left heatmap \new{(\textbf{T1} and \textbf{T2})} shows seating positions across multiple metrics, allowing rapid assessment of relative performance by play order.
Finally, an icicle plot \new{(\textbf{T4})} at the bottom-right encodes the number of decks per color identities and number of colors with an extra brightness encoding for player ``saltiness'', highlighting patterns in player experiences and specific color combinations and commanders. \new{We chose this type of chart because it performs well compared to hierarchical alternatives such as treemaps \cite{Macquisten2020-in}.}
Finally, the dashboard includes panels for active filters and customization instructions, ensuring that users can tailor views to specific analytical questions and comparative contexts. \new{Our next step was to evaluate the dashboard. We retained it as a static, black-and-white prototype, since research indicates that prototype medium has no significant main effect on users' ability to detect usability problems or on their perceptions of usability \cite{Boothe2013-zx}.}

\subsubsection{Evaluation}
We met with the Playgroup.gg team to gather their feedback on the proposed dashboard.
While they praised the overall design, they noted one weakness: using only a black stroke for the marks in the line chart made it difficult to track a single line across the x-axis.
After consideration, we decided to retain the black stroke, as the goal of the visualization was not to individually identify commanders since this functionality was already supported through filtering in the right-hand panel.
Instead, we want to analyze the overall evolution of damage per turn.
Following this, we collaborated with the Playgroup.gg team to develop a questionnaire designed to assess whether users could understand and extract insights from the charts.
The questionnaire contained one section per chart, with each item structured as a multiple-choice question with only one correct answer.
Each item required participants to interpret a chart and respond accordingly.
To reduce random guessing, participants were explicitly instructed not to guess and were given the option to select \textit{I don't know} if undecided.
We framed the questionnaire as a quiz, awarding one point per correct answer, to foster engagement.
Finally, participants were invited to provide open-ended feedback as well as demographic information to contextualize the sample.

\subsubsection{Results}

Released via Playgroup.gg's website, Reddit, and Discord, 83 participants (77 were current Playgroup.gg users; 92.77\%) completed the 33-item questionnaire. Participants demonstrated relatively strong comprehension with an average raw score of 25.05 (SD = 6.22), ranging from 6 to 33. The 95\% CI [23.69, 26.41] suggests correct answers on approximately three-quarters of items. Corrected scores (penalizing guessing) showed similar patterns: mean 23.97 (SD = 6.59), range 4.85–33.

Performance varied by chart type. Heatmaps yielded the highest accuracy (98.19\%), with nearly all identifying seating position outcomes correctly. Parallel coordinates showed high comprehension (91.57\%), indicating effective relative comparison extraction. Line charts performed strongly (78.92\%), suggesting successful temporal pattern interpretation despite earlier distinguishability concerns. Bar charts averaged 76.81\%; dominant category identification reached 100\%, but comparative reasoning dropped below 50\%. Scatterplots proved challenging (67.47\%); while high-level features like most-played commander (80\%) or specific game counts (87\%) were easily identified, precise winrate interpretation fell below 50\%. Icicle plots showed 75.30\% overall; structural features exceeded 75\% accuracy, but individual commander pinpointing dropped to 49\%. Thus, scatterplots and icicle plots supported broad pattern recognition but demanded higher cognitive effort for precise lookups.

Participant feedback consistently highlighted difficulty interpreting black-and-white visuals. Repeatedly, "colors would make these much easier to read" was cited—especially for overlapping line graphs. Requests included colorblind-friendly palettes and user-controlled coloring. Individual chart reactions were mixed: heatmaps and bars were praised as "clear"; scatterplots drew skepticism regarding shade encoding ("might be better with circle size"); and lines confused without color. Icicles received the harshest criticism ("eye-sore," "gave me a headache"), though some suggested boundary strengthening or interactive callouts.

Interactive feature requests emerged prominently: hovering/clicking for precise values, filtering by power level/game length, and brushing/selection across linked charts. Customization was similarly recurrent—"remove/add charts," change graph states, or select preferred chart types. Many noted excessive information density and intimidation when all displays appeared simultaneously.\new{Participants thus suggested breaking into sections, displaying one-at-a-time, or quick-glance summaries before drilling detail, attributing concerns partly to moderate-to-low adult data visualization literacy varying across chart types \cite{Galesic2010}.} Users requested clearer chart separation and consistent legend placement.

Additional proposals included: (i) commander images/icons, (ii) direct labels/percentages on marks, (iii) missing metrics (average turn time, power level filters), and (iv) light/dark modes. Despite criticisms, enthusiasm persisted—the dashboard called "very interesting," "good addition," and "great information." Some noted usefulness increases after familiarity, acknowledging potential learning curves alongside eventual value.
\section{Analysis}

The final design study phase consists of reflect and write stages; the latter produced this paper. Reflect encompasses all preceding stages except implementation/deployment.

Our results confirm, refine, and propose guidelines for gameplay analysis dashboards in platforms like Playgroup.gg. First, we confirm existing guidance on contextually relevant, outcome-driven metrics: players valued win rate and wins/losses while rejecting peripheral measures (dice rolls, pause duration). Second, we refine visualization choice guidelines: canonical charts (heatmaps, parallel coordinates, line charts) showed high comprehension (M higher or equal to 78.9), but scatterplots and icicle plots posed challenges under non-ideal conditions (e.g., black-and-white rendering). We recommend pairing complex/unconventional charts with interactivity (hover-to-reveal, filtering, brushing) or simplified encodings. Third, we reject the assumption that global/population-level dashboards are inherently useful: participants strongly preferred playgroup-/user-level views over aggregated data. Fourth, we propose new guidelines emphasizing customization and progressive disclosure. Participants requested tailored chart selection, preferred encodings, and controlled density (quick summaries before detail)—underscoring that adaptability is as critical as accuracy. Together, these reflections extend design study knowledge by grounding visualization guidelines in an illustrative community's practices while pointing to generalizable principles for interpretability, customization, and contextual relevance.

In sum, we successfully applied Sedlmair et al.'s \cite{sedlmair2012design} design study methodology—a \ac{PD} extension \cite{frishberg2011interactive,kahng2017cti}—to \ac{MTG}, contributing to \ac{InfoVis} by introducing a \ac{TCG} previously unexplored in data visualization prototypes \cite{Wallner2014}. While \ac{MTG} has been explored by the community \cite{githubGitHubVincenzoGalantemagicthegathering}, sound research methodologies focused on gameplay were lacking—which we addressed. By documenting stage details from prior sections, we hope the \ac{InfoVis} community can adopt our use case as an example for other domains (\ac{TCG}s or game types). Though we followed user needs for dashboard proposals, we also hope to spark curiosity about new visual techniques, demonstrating patterns many remain unaware of \cite{Goyal2013} due to low visualization literacy \cite{Galesic2010}, while showcasing collected data presented creatively for Playgroup.gg creators through \ac{InfoVis}.
\new{\section{Future Work}}
The proposed design should be reviewed in light of our results and continue with a prototype implementation and deployment in the Playgroup.gg website.
Additionally, researchers may further investigate how interactive features can mitigate the challenges observed with complex visualizations such as scatterplots and icicle plots, particularly in mobile or constrained environments where overlapping elements are harder to trace. 
Further studies could explore adaptive dashboards that personalize chart selection and level of detail to different user profiles, such as novice players seeking simple summaries versus expert players interested in advanced cross-metric analyses.
Incorporating player-centered enhancements such as commander images, mana symbols, or icons may improve recognition and engagement, and systematic testing of accessibility features (e.g., colorblind palettes, light/dark modes) could yield generalizable guidelines. 
Importantly, future research should broaden the participant base beyond Playgroup.gg users. 
Since our sample primarily consisted of individuals already interested in statistics and data tracking, their preferences may not fully represent the general Commander audience.
Comparing results across both data-inclined and casual player populations would clarify whether the observed prioritization of strategic, outcome-driven metrics reflects a universal trend or a community-specific bias.
Finally, longitudinal research is needed to understand how player preferences and comprehension evolve over time as familiarity with the dashboard grows, helping to refine guidelines around learning curves and long-term adoption.
\section{Conclusions}
This design study investigated the design and evaluation of a dashboard for analyzing gameplay data in \ac{MTG} Commander, implemented in collaboration with the Playgroup.gg platform.
Through iterative design, user feedback, and large-scale questionnaire testing, we examined which metrics, analyses, and visualization types players found most valuable, and how effectively they could interpret the resulting dashboard.
Overall, this study contributes empirical evidence and practical guidelines for designing dashboards in gameplay analysis contexts, while also informing broader visualization research on user-centered design, comprehension, and engagement.

\section*{Acknowledgment}
The authors wish to thank Maran and Johannes for the opportunity. Work supported by national funds through Fundação para a Ciência e a Tecnologia, I.P. (FCT) under projects UID/6486/2025, UID/PRR/6486/2025, UID/PRR2/06486/2025, UID/50021/2025, UID/PRR/50021/2025, and UID/315/2025.

\bibliographystyle{IEEEtran}
\bibliography{papers}

\end{document}